\documentclass[fleqn,usenatbib,useAMS]{mnras}

\usepackage{graphicx}		% Including figure files
\usepackage{amsmath}		% Advanced maths commands
\usepackage{amssymb}		% Extra maths symbols
\usepackage{multicol}        	% Multi-column entries in tables
\usepackage{bm}			% Bold maths symbols, including upright Greek
\usepackage{pdflscape}		% Landscape pages

\usepackage[T1]{fontenc}
\usepackage{ae,aecompl}

\usepackage{newtxtext,newtxmath}

\defcitealias{gre13}{G13}

\title[Halo kinematics among carbon-rich dwarfs]{Evidence for halo kinematics among cool carbon-rich dwarfs}

\author[J. Farihi et al.]{J. Farihi$^{1}$\thanks{E-mail: j.farihi@ucl.ac.uk},
A. R. Arendt$^{1}$,
H. S. Machado$^{2}$,
L. J. Whitehouse$^{1}$
\\
$^{1}$Department of Physics and Astronomy, University College London, London, UK\\
$^{2}$Department of Astronomy, University of California, Berkeley, USA
}

\begin{document}

%\label{firstpage}

\maketitle

\begin{abstract}
This paper reports preliminary yet compelling kinematical inferences for $N\ga600$ carbon-rich dwarf stars that demonstrate 
around 30\% to 60\% are members of the Galactic halo.  The study uses a spectroscopically and non-kinematically selected 
sample of stars from the SDSS, and cross-correlates these data with three proper motion catalogs based on {\em Gaia} DR1 
astrometry to generate estimates of their 3-D space velocities.  The fraction of stars with halo-like kinematics is roughly 30\%
for distances based on a limited number of parallax measurements, with the remainder dominated by the thick disk, but close
to 60\% of the sample lie below an old, metal-poor disk isochrone in reduced proper motion.  An ancient population is consistent 
with an extrinsic origin for C/O $>1$ in cool dwarfs, where a fixed mass of carbon pollution more readily surmounts lower oxygen 
abundances, and with a lack of detectable ultraviolet-blue flux from younger white dwarf companions.  For an initial stellar mass 
function that favors low-mass stars as in the Galactic disk, the dC stars are likely to be the dominant source of carbon-enhanced, 
metal-poor stars in the Galaxy.

%{\em 
%Gaia} should confirm this preliminary finding and determine the precise fraction of halo and thick disk systems, as well as 
%any carbon-rich dwarfs with thin disk origins.

\end{abstract}


\section{Introduction}

The first dwarf carbon star was confirmed over 40 years ago, and, as with many astronomical discoveries past and present, 
it was discovered serendipitously in a large survey.  G77-61 was first identified in the Lowell Observatory proper motion survey 
\citep{gic61}, with the same purpose-built telescope and data that successfully discovered Pluto \citep{gic58}.  Some decades 
later when a trigonometric parallax was obtained by the US Naval Observatory, the cool main-sequence star was initially 
unremarkable, and only later noted to be unusually red for its absolute magnitude.  An optical spectrum eventually revealed 
strong molecular carbon bands, making G77-61 the first dwarf carbon (dC) star \citep{dah77}.

During the intervening decades, scientific progress on the origin and broader Galactic context of dC stars has been modest.  
Prior to the Sloan Digital Sky Survey (SDSS) roughly one dozen dC stars were identified by proper motion and objective prism 
surveys \citep{bot91,gre91,war93,heb93,lie94,low03}, for which a giant stellar luminosity and thus distance \citep{wal98} would 
imply highly unbound Galactic velocities.  To date, only three dC stars have published trigonometric parallaxes \citep{har98}, 
all with 9.6\,mag $\leq M_V\leq10.0$\,mag, while ongoing work at the USNO indicates this range can vary up to around $\pm2
$\,mag \citep{pla16}.  Until forthcoming {\em Gaia} data releases that include a substantial number of reliable dC star parallaxes
(duplicity may influence measurement errors) proper motion remains the best universal indicator of carbon star luminosity.

The SDSS has been transformative in terms of increasing both the number and diversity of definite and candidate dC stars. In 
the early years of the survey, including commissioning and DR1 data, spectroscopic fibers were dedicated to search for carbon 
stars on the basis of their photometric colors.  This yielded a few hundred objects and increased the number of possible dC 
stars by roughly an order of magnitude \citep{mar02,dow04}.  Proper motions based on the USNO-B1 catalog \citep{mon03} 
suggested the majority of the few hundred carbon stars thus discovered were bona fide dwarfs.  However, by far the largest 
spectroscopic sample of (faint) carbon stars identified to date is the prodigious collection of 1211 SDSS sources described 
by \citet[][hereinafter G13]{gre13}.

The prototype dC star has both clear halo kinematics and is among the most metal-poor stars known \citep{dah77,ple05}.  
In addition, there are a handful of proper-motion selected dC stars with parallaxes and high velocity that are either halo or 
thick disk members \citep{war93,har98}.  On the other hand, there are two spectroscopic composite dC systems containing 
a sufficiently blue white dwarf that were known prior to SDSS \citep{heb93,lie94}, and another nine systems have been found 
in SDSS spectra (\citealt{reb10}; \citetalias{gre13}; \citealt{si14}).  These systems appear to have disk kinematics and this is 
consistent with the relative warmth and youth of the white dwarf components.  A thorough and unbiased investigation into the 
Galactic origin of all dC stars is likely to require both {\em Gaia} and the large ground-based spectroscopic follow-up surveys 
coming online in the next few years.

In this work, the SDSS carbon star sample of \citetalias{gre13} is cross-matched with recently published proper motion catalogs 
based on astrometry from {\em Gaia} DR1, and SEGUE pipeline radial velocities, yielding robust 3-D data for over 500 to 600 
sources.  These astrometric datasets represent several substantial improvements over previously available measurements,
especially for faint and both fast- and slow-moving sources \citep{mun04}.  Reduced proper motions for a sub-sample of over 
600 stars with confident measurements suggests the bulk of dC stars lie below the thin-disk main sequence, and contain both 
thick-disk and halo stars of sub-solar metallicities.  Based on existing parallax data, $UVW$ space velocities are calculated for
known dC star absolute magnitudes, where a median luminosity yields a fraction of halo stars that is at least 30\%, and a clear 
thick disk component.  The data, catalog matching, and sub-sample selection is described in Section 2, and the results are 
presented in Section 3.  Preliminary conclusions are presented in Section 4 with an eye toward confirmation with {\em Gaia} 
DR2.

\section{Sample Definition and Selection}

This study makes use of existing SDSS catalog data, correlating a large sample of dC star candidates with both radial velocity 
determinations and with state-of-the-art proper motion catalogs based on {\em Gaia} DR1.  Below are described the overall
sample and catalogs, and additional criteria imposed to ensure a relatively high-fidelity sub-sample for analysis.

\subsection{Catalog Data}

The basis of this kinematical study is the catalog of carbon stars from \citetalias{gre13}, where selection was not based on proper 
motion or photometric color, but instead took advantage of cross-correlation between existing spectroscopic targets and carbon 
star spectral templates within SDSS DR7 and DR8.  In addition to a modest number of spurious objects that are readily rejected 
by visual inspection, \citetalias{gre13} also identified dozens of carbon-rich white dwarfs, as well as a similar number of bluer 
stars showing carbon bands (referred to as G-type) that strongly cluster in color space and which may be warmer counterparts to 
dC stars.  While the sample may be incomplete or insensitive to dC stars with spectra that are sufficiently distinct from classical 
carbon giants, there is no {\em kinematical} bias, and hence the space motions of hundreds of carbon stars can be analyzed 
with confidence.

The \citetalias{gre13} sample of dC candidates was cross-matched with three recently published proper motion catalogs, all 
of which are based on {\em Gaia} DR1 astrometry.  The first catalog is that employed by \citet{dea17}, where {\em Gaia} DR1 is 
used to recalibrate the SDSS astrometric mapping from first principles.  Because this catalog was produced and provided by the 
Institute of Astronomy (V. Belokurov 2017, private communication), it is referred to as IoA hereinafter.  The second proper motion 
catalog covering the SDSS footprint is HSOY \citep{alt17}, and based on combining {\em Gaia} DR1 astrometry with PPXML 
\citep{roe10}, where the latter is tied to astrometry from 2MASS and USNO-B1.  The third catalog is GPS1 \citep{tia17} and 
is a combination of astrometry from {\em Gaia} DR1, Pan-STARRS, and SDSS.

Proper motion tables were uploaded locally and from VizieR\footnote{http://vizier.u-strasbg.fr/viz-bin/VizieR} into 
{\sc topcat}\footnote{http://www.star.bris.ac.uk/$\sim$mbt/topcat} for manipulation.  It was first necessary to cross-match the 
J2000 SDSS designations from \citetalias{gre13} Table 1 with the DR12 database to obtain up to date SDSS parameters including, 
critically, {\sf objID} and {\sf specObjID}, which only became unique and permanent catalog identifiers after DR7.  Cross-matching 
with the IoA proper motion catalog was done by {\sf objID}, whereas for HSOY and GPS1 the catalog matching was done by right 
ascension and declination converted to degrees from J2000 designations.  In these latter two cases, a search radius of 6\arcsec 
\,was found to be ideal, as the epochs of some GPS1 catalog positions were ambiguous.  The matched tables were then joined 
using concatenation in {\sc topcat}.

A critical part of the available kinematical data for the carbon star sample are the radial velocities sourced from the SEGUE
survey and pipeline \citep{lee08,yan09}.  The radial velocities thus produced were measured by cross-correlation against 
either the SDSS commissioning templates \citep{sto02}, or ELODIE spectroscopic templates \citep{pru01} degraded to 
SDSS resolution, with ELODIE templates often providing the best match.  Each entry in the cross-matched catalogs above 
was queried with the {\sf crossid} software provided by the SDSS, by matching its position to the nearest primary object, 
with the retrieved primary {\sf objID} checked to ensure it matched that of each dC candidate.  

For the 1211 objects in Table 1 of \citetalias{gre13}, the above SDSS DR12 query produced a total of 1344 matches, thus 
indicating that 112 dC candidates possessed two or more spectroscopic observations, and therefore the same number of 
{\sf specObjID}.  These objects with duplicate spectra were flagged, evaluated, and consolidated so that all cross-matched 
catalogs had exactly 1211 entries.  Stars with more than a single spectrum and radial velocity determination were assessed 
by taking the mean radial velocity, after weighting by the spectroscopic S/N.  

\subsection{Data Cuts}

\begin{figure}
\includegraphics[width=84mm]{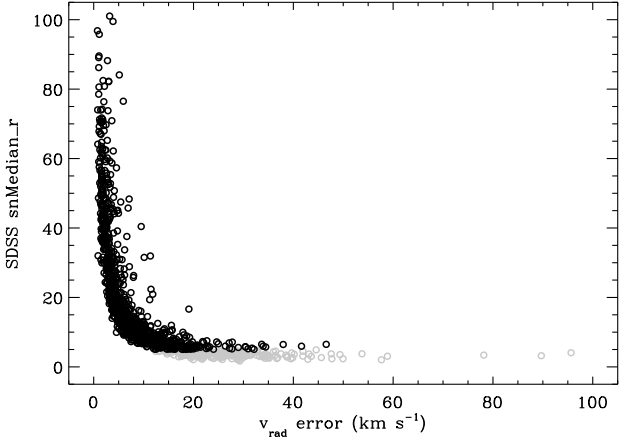}
\caption{SDSS and SEGUE pipeline parameters for the error in radial velocity from spectral template matching, versus the 
spectroscopic S/N in the $r$-band.  This latter quantity, {\sf snMedian\_r} was in general the highest of all the available S/N 
diagnostics in both the standard SDSS and SEGUE spectroscopic parameters.  There are 1187 stars with determined radial 
velocities plotted as grey circles, but after a cut at {\sf snMedian\_r} $> 5$ to eliminate unreliable data, there are 961 stars 
remaining and overplotted as black circles.
\label{fig1}}
\end{figure}

Figure \ref{fig1} plots the SEGUE pipeline reported error in radial velocity as a function of the SDSS pipeline parameter {\sf 
snMedian\_r} (spectroscopic S/N over the photometric $r$ band), demonstrating a clear correlation.  After visual inspection 
of all 1211 spectra, only sources with {\sf snMedian\_r} $>5$ were retained for analysis.  This data cut served not only to 
remove objects with large uncertainties in their radial velocity, but also to ensure the target spectrum was that of a genuine 
carbon star, as a handful of contaminants were identified by visual inspection \citepalias[see][]{gre13}.

%%%FIGURE 1%%%

Another requirement for the kinematical analysis was that the total proper motion be greater than at least a few standard errors.  
These vary among the three proper motion catalogs, and are generally constant near 2\,mas\,yr$^{-1}$ up to 18$^{\rm th}$ 
magnitude, but can rise to 4\,mas\,yr$^{-1}$ at the faint end of SDSS $r$ band or {\em Gaia} $G$ band \citep{dea17,alt17,tia17}. 
Because the \citetalias{gre13} sample contains over 600 sources fainter than $r=19.0$\,AB\,mag, the adopted requirement is a 
total proper motion $\upmu > 10$\,mas\,yr$^{-1}$, which should be at least $5\upsigma$ for brighter stars and always above 
$2\upsigma$.

The top of Table \ref{tbl1} lists the number of \citetalias{gre13} stars present in various proper motion catalogs, and those 
remaining after the two data cuts.  It is noteworthy that all three proper motion catalogs provide a similar total number of raw 
matches with the dC star candidate sample, and also after the proper motion cut, but then differ by several tens of stars after
the S/N requirement is imposed.  In the highest fidelity sub-samples, the IoA catalog contains the fewest matches, which appears
to be due to a deficit of the highest proper motion stars.  Relative to GPS1, the IoA catalog has 34 fewer sources with $\upmu > 
100$\,mas\,yr$^{-1}$, and 55 fewer sources with $\upmu > 50$\,mas\,yr$^{-1}$.  Owing to the highest number of reliable matches
(and thus likely completeness) in GPS1, results from this catalog are adopted here, but the findings from all three catalogs are 
similar and listed in Table \ref{tbl1}.

\section{Kinematical Inferences}

In the following, both reduced proper motion and 3-D space velocity estimates are used to infer the likely Galactic orbital and 
birth regions of bona fide dC stars.  Galactic $UVW$ space velocities are calculated for all three sets of proper motion catalog 
cross-matches, and for a range of distance estimates based on measured dC star parallaxes.  

\subsection{Reduced Proper Motion}

\begin{figure}
\includegraphics[width=84mm]{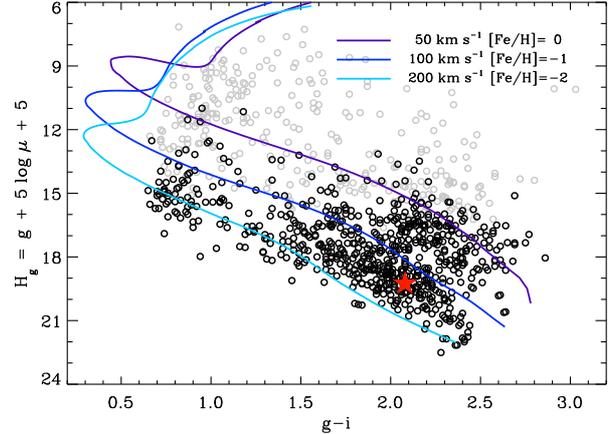}
\caption{Reduced proper motion in $g$ band versus $g-i$ color, both in AB magnitudes, using the GPS1 cross-matched sample 
of 980 stars plotted as grey circles.  After making a total proper motion cut $\upmu > 10$\,mas\,yr$^{-1}$ to eliminate sources with 
proper motion consistent with zero (see text), this number reduces to 677 stars that are plotted as black circles.  Three stellar 
isochrones are plotted using the Dartmouth stellar evolution models \citep{dot08}, and are intended to serve as rough guides 
for the (older) thin disk, thick disk, and halo.  The position of G77-61 is marked with a large red star, and while it is known to be 
a halo star and one of the most metal-poor objects known, its position suggests it is not far from the norm.
\label{fig2}}
\end{figure}

In this section, results for the GPS1 cross-matched sub-sample are analyzed, and in the next section all three catalogs are
compared (and shown to be similar).  Reduced proper motions were calculated following \citet{mun17}, using their Equation 2.  
Ignoring corrections for Galactic latitude, the reduced proper motion in SDSS $g$ band is
\begin{equation}
H_g = g + 5\log{\upmu} + 5 = M_g + 5\log{v_{\rm tan}} - 3.379
\end{equation}
where $\upmu$ is the magnitude of the proper motion in arcsec\,yr$^{-1}$ and $v_{\rm tan}$ is the velocity in the plane of the sky 
in km\,s$^{-1}$.  The left-hand side of the equation is used to plot the GPS1 crossed-matched stars in Figure \ref{fig2}, and the
right-hand side is used to plot theoretical isochrones.  First are plotted all 980 stars in common with \citetalias{gre13} and GPS1, 
and then this number is reduced to 677 stars with significant proper motions.  The plot demonstrates that the cut to eliminate
proper motions consistent with zero generally acts to de-populate the brighter regions of the diagram.  The first noteworthy feature 
of Figure \ref{fig2} is the possibility of up to three weakly defined loci.  There is a dense group radiating along ($x,y$) = (2.0, 19), a 
more diffuse clump near (2.2, 17), and a relatively small number of stars focussed around (0.8, 15).  If these features are real,
which is not completely certain, then they could hint at clusters of stars with similar metallicity and kinematics.  The latter group 
with the bluest colors are almost certainly the G-type carbon dwarfs described by \citetalias{gre13}.

%%%FIGURE 2%%%

Three representative stellar isochrones are overplotted on Figure \ref{fig2}, and are taken from the Dartmouth stellar evolution
database \citep{dot08}.  The choice of isochrones is somewhat arbitrary, and they were selected to represent a range of ages,
metallicities, and associated kinematics as a visual guide to the broad characteristics of the reduced proper motions.  It should
also be noted that for a given stellar mass and $T_{\rm eff}$, carbon dwarf colors should be somewhat distinct from those of
their oxygen-rich, K and M dwarf counterparts.  Model stellar atmospheres and spectra are not yet developed and available for
dC stars, but these isochrones should nevertheless be instructive.  The top track is for solar metallicity stars of age 5\,Gyr, with 
tangential speed 50\,km\,s$^{-1}$ and should be a good proxy for older objects of the thin disk.  The middle model is for modestly 
metal-poor stars of age 10\,Gyr, moving at 100\,km\,s$^{-1}$, and should be a decent approximation for the metal-poor thick 
disk \citep{fuh04}.  The bottom isochrone is for metal-poor Population II stars of age 15\,Gyr and speed 200\,km\,s$^{-1}$, 
as would be expected for halo denizens.  None of the isochrones have $\alpha$ element abundances different from solar.

While none of these definitively overlap with densely plotted regions of the diagram, they do provide a general indication of the
underlying population.  Ignoring the stars rejected due to proper motions consistent with zero, it is clear {\em these are all dwarf 
stars}.  That is, based on the GPS1 measurements, there are at least 677 bona fide dC stars confirmed by their combination of
brightness and proper motion.  Another tentative conclusion that may be drawn is that there appears to be only a small fraction 
of thin disk stars, and the bulk of dC stars are more consistent with thick disk and halo origins.  About 92\% of the plotted black
sources lie below the isochrone representing the thin disk, and nearly 60\% lie below the isochrone depicting thick disk stars.
Interestingly, G77-61 sits near the main locus and above the halo-like isochrone in Figure \ref{fig2} even though it is known to
be an extremely metal poor star.  Thus the diagram is likely useful for the broad characteristics of the dC population, but likely 
not for any individual stars.

\subsection{Galactic Space Velocities}

The 3-D space velocities were calculated for all three proper motion catalog subsamples to ensure the results were 
consistent.  Here, ($U,V,W$) refers to a right-handed coordinate system where $U$ is positive toward the Galactic anti-center, 
$V$ is positive in the direction of Galactic rotation, and $W$ is positive towards the North Galactic pole.  The sample of SDSS 
dC stars does not have parallaxes or other distance determinations, and thus the velocities in the plane of the sky depend 
linearly on the actual distances.  Below, a range is explored that should encompass the bulk of dC stars.

While there are only three dC stars with published parallaxes, roughly 20 stars have been monitored by the USNO over the 
past decades.  \citet{har98} report 9.6\,mag $\leq M_V \leq 10.0$\,mag for three stars, with G77-61 at the faint end of this range, 
and \citet{pla16} cite unpublished work where dC stars measured to date have 7.5\,AB\,mag $< M_r < 11.5$\,AB\,mag (note the 
use of the SDSS $r$ band rather than $V$).  Translating the $r=13.2$\,AB\,mag \citep{zac13} measured for G77-61 into an 
absolute magnitude yields $M_r=9.4$\,AB\,mag for this prototype dC star, and this places it in the center of the absolute 
brightness distribution.  

Based on the above parallax data, and the fact that G77-61 is a part of the principal, broad locus in the reduced proper motion 
diagram, it is likely that $M_r=9.5$\,AB\,mag is a good approximation for the bulk of dC stars.  However, to ensure that that 
any inferences are not skewed by an inappropriate choice of stellar distances, $UVW$ calculations are also made for absolute 
magnitudes $\pm1.5$\,mag times brighter and fainter.  Ideally, a better method could employ a random sampling of $M_r$ 
based on a known distribution, or use dC star colors as a proxy for $T_{\rm eff}$ and hence luminosity.  However, neither of 
these approaches is currently feasible.  The distribution of dC star luminosities is not well-determined, and the USNO program 
stars were intentionally selected to sample a range of properties and should not be considered representative of the underlying 
population.  Despite the range of $M_r$ quoted above, the colors of stars with parallaxes are tightly clustered (J. A. Munn 2018, 
private communication).  For this reason, and because mass, metallicity, and carbon abundance all affect luminosity and color, 
using a range of $M_r$ for the entire sample is the most efficient approach prior to {\em Gaia} parallax measurements.

%%%FIGURE 3%%%
\begin{figure*}
\includegraphics[width=172mm]{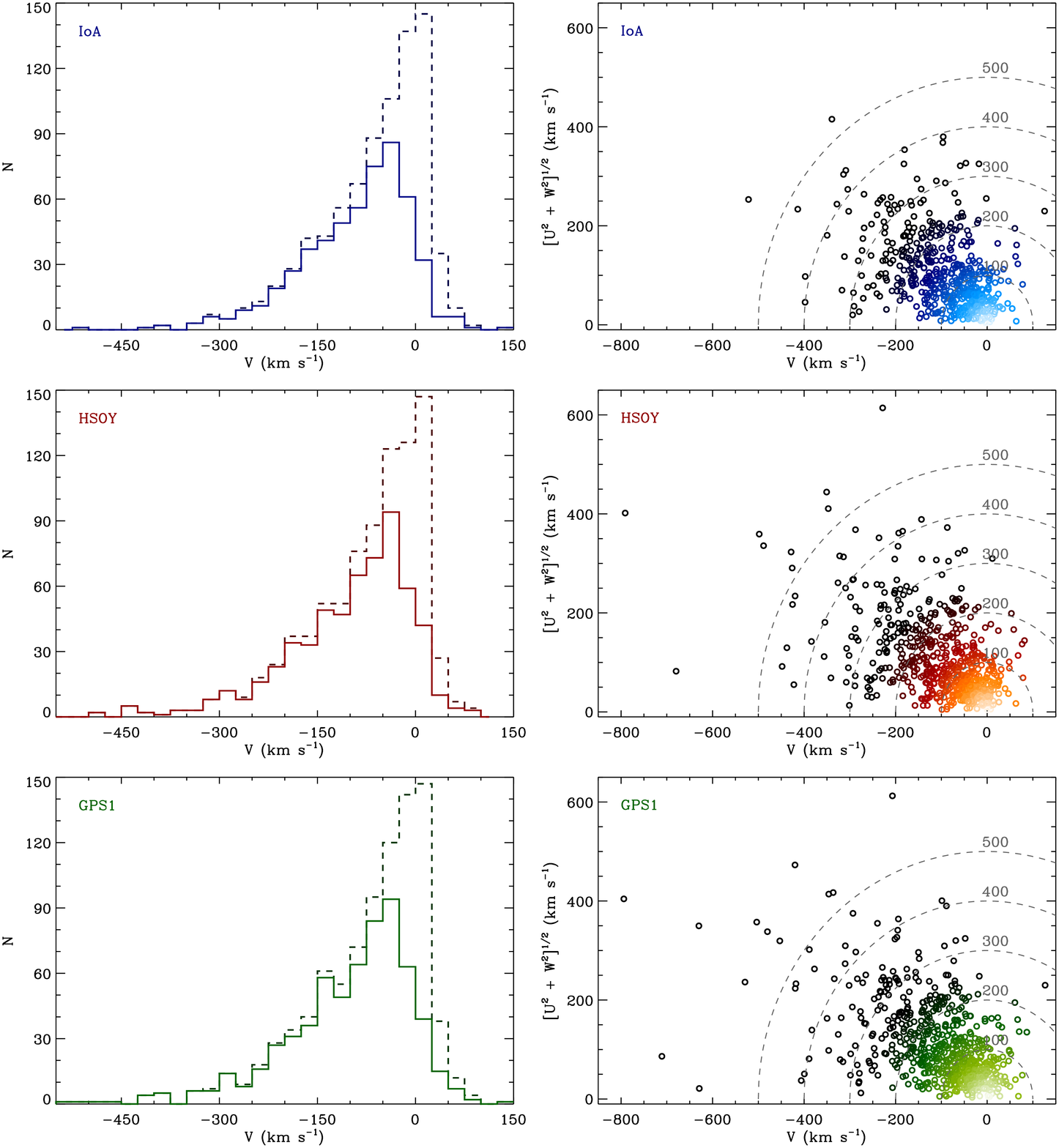}
\caption{Space velocity plots for the sub-samples of SDSS dC candidates with matches in each of the IoA, HSOY, and GPS1 
proper motion catalogs, with $M_r=9.5$\,AB\,mag.  The left-hand panels are Galactic $V$ velocity histograms.  The dashed lines 
show the number of catalog matches that also have an SDSS spectroscopic pipeline parameter {\sf snMedian\_r} $>5$.  The solid 
lines includes only those subsets with total proper motion $\upmu>10$\,mas\,yr$^{-1}$, and thus measurements that are greater 
than $2\upsigma$ to $5\upsigma$ \citep{dea17,alt17,tia17}.  Following on the exclusion of proper motions consistent with zero, 
the peaks in $V$ all shift to negative velocities that lag behind Galactic disk rotation (see Table \ref{tbl1}).  The right-hand panels 
are Toomre diagrams for the solid-lined histogram stars.  Contours for total (peculiar) space velocity $T^2=U^2 + V^2 + W^2$ are 
labelled as dashed grey lines, and the color of each data point is weighted by $T$ with white corresponding to zero and black to 
250\,km\,s$^{-1}$ and beyond.  There are almost 200 stars with unambiguous, halo-like velocities among the stars in the GPS1 
sub-sample, which has the highest number of matches.  However, with $T=209$\,km\,s$^{-1}$, G77-61 would not stand out in 
these diagrams, and so the halo fraction estimates in Table \ref{tbl1} are likely to be somewhat conservative.
\label{fig3}}
\end{figure*}

Figure \ref{fig3} plots all three sub-samples of dC stars as Galactic $V$ histograms, and as Toomre energy diagrams in order to
distinguish disk- and halo-like kinematics.  While these are all essentially similar, the absence of the relatively high proper motion 
stars in the IoA catalog is apparent.  Two broad conclusions can be tentatively drawn from these results, on the basis that $M_r=
9.5$\,AB\,mag is a decent approximation of the dC population.  First, dC stars lag behind the rotation of the Galactic disk, implying 
that old stars are the dominant species, either thick disk or halo.  The non-Gaussian shape of the distribution indicates there are 
at least two components.  Second, there are a substantial number of stars with total (peculiar) space velocities above 200\,km\,s$
^{-1}$ and that are therefore halo members.  This may be a conservative estimate, as halo stars are often have total velocities 
above $150-200$\,km\,s$^{-1}$ \citep{pau06}, and the reduced proper motion diagram suggests a higher halo fraction.

Table \ref{tbl1} lists the statistical characteristics for each of the three proper motion catalog subsamples, and for each of 
three absolute magnitude and hence distance approximations.  For each of the latter, the mean Galactic $V$ is listed together 
with its dispersion and two estimates of the fraction of stars with halo-like space motions.  Stars with $V<-200$\,km\,s$^{-1}$ 
are not rotating with the Galactic disk, and because there should be an equal number of retro- and pro-grade halo objects,
this fraction is doubled to approximate a Gaussian distribution of non-rotating stars.  Another statistic selects stars with $T> 
200$\,km\,s$^{-1}$ as objects with probable halo velocities.  These fractions may be lower limits to the true fraction of halo 
stars, as their distribution in $V$ is only a single dimension \citep{chi00,fuh04}, and total space motion ignores directionality 
(and dispersion).

 %%%TABLE1%%%
\begin{table}
\begin{center}
\caption{Space velocity inferences for a sub-sample of bona fide dC stars.\label{tbl1}}
\begin{tabular}{@{}rrrr@{}}

\hline

Catalog									&IoA$^{\rm a}$		&HSOY			&GPS1\\
\hline

$N$(G13 $\cap$ SDSS $v_{\rm rad}$)			&978				&962				&965\\
$N$($\cap$ $\upmu > 10$\,mas\,yr$^{-1}$)		&652				&680				&671\\
$N$($\cap$ {\sf snMedian\_r} $>5$)				&535				&598				&637\\
         
\hline

\multicolumn{4}{c}{$M_r = 11.0$\,AB mag:}\\
\\ 
$\langle V \rangle$ (km\,s$^{-1}$)				&--53			&--57			&--58\\
$\upsigma_V$ (km\,s$^{-1}$)					&61				&68				&69\\
$2\times f(V<-200$\,km\,s$^{-1}$)				&0.04			&0.08			&0.08\\
$f(T>200$\,km\,s$^{-1}$)						&0.14			&0.16			&0.16\\      

\hline

\multicolumn{4}{c}{$M_r = 9.5$\,AB mag:}\\
\\ 
$\langle V \rangle$ (km\,s$^{-1}$)				&--93			&--103			&--105\\
$\upsigma_V$ (km\,s$^{-1}$)					&85				&102				&107\\
$2\times f(V<-200$\,km\,s$^{-1}$)				&0.21			&0.28			&0.30\\
$f(T>200$\,km\,s$^{-1}$)						&0.28			&0.30			&0.32\\     

\hline

\multicolumn{4}{c}{$M_r = 8.0$\,AB mag:}\\
\\ 
$\langle V \rangle$ (km\,s$^{-1}$)				&--174			&--195			&--197\\
$\upsigma_V$ (km\,s$^{-1}$)					&146				&182				&195\\
$2\times f(V<-200$\,km\,s$^{-1}$)				&0.71			&0.77			&0.78\\
$f(T>200$\,km\,s$^{-1}$)						&0.54			&0.56			&0.56\\  

\hline
     
\end{tabular}
\end{center}

\flushleft

\smallskip
$^{\rm a}$ This catalog appears incomplete for total proper motions $\upmu > 50-100$\,mas\,yr$^{-1}$ (see text).\\

\end{table}

The most conservative result is that calculated for $M_r=11.0$\,AB\,mag and suggests that at least around 10\% to15\% 
are halo stars, and that a typical dC star lags behind the Galactic disk by $50-60$\,km\,s$^{-1}$.  In this case there would be 
a substantial old disk component in the stellar population, because the thick disk has $\langle V\rangle=-30$\,km\,s$^{-1}$ 
\citep{chi00}. The least conservative estimates are calculated for $M_r=8.0$\,AB\,mag and would imply that at least 55\% of 
dC stars are members of the halo; while potentially remarkable, it is unlikely based on parallax work that this is the case (and 
the dispersion in $V$ would be roughly twice that of halo stars!).  The case adopted here for interpretation is the middle panel 
of the Table for $M_r=9.5$\,AB\,mag, where at least about 30\% of dC stars were formed in the Galactic halo.  In this case 
the dC population has a $\upsigma_V$ similar to halo stars, but where the mean is roughly half that for a pure halo population.  
This is probably best explained by a mixture of halo and thick disk stars.  It is noteworthy that the middle luminosity assumption 
produces a similar fraction of stars with halo-like kinematics via two diagnostics, and also consistent with the reduced proper 
motions.

\section{Conclusions}

With the largest sample of dC star candidates assembled to date, and proper motion motion catalogs derived from {\em Gaia}
DR1 astrometry, the analysis here demonstrates that at least around 680 are genuine main-sequence stars, and that a significant 
fraction have kinematical properties consistent with halo membership.  Importantly, this spectroscopically identified sample is 
free of kinematical bias towards older and higher velocity stars.  

The exact fraction of halo stars among carbon-rich dwarfs depends on the distance estimates used here based on a handful 
of parallaxes, and also where halo objects are distinguished from high velocity disk stars.  If objects with total space velocity 
below 200\,km\,s$^{-1}$ are considered to be disk stars, then the halo fraction is around 30\% for a population with $M_r\approx
9.5$\,AB\,mag.  The distribution of Galactic $V$ velocities suggests that dC stars are primarily a combination of halo and old thick 
disk stars, with a small but real thin disk component.  Overall, the dC stars lag behind the rotation of the Galaxy and are thus 
consistent with a relatively old population.  Their positions in a reduced proper motion diagram suggest the bulk of stars are metal 
poor, and a halo fraction as high as 60\%.  All of the above is consistent with the lack of photometric and spectral evidence for warm 
to hot white dwarf companions to the bulk of dC stars.

In a scenario where dC stars are formed via binary mass transfer from a carbon-rich giant \citep{dah77}, stars with an intrinsically 
lower metal -- and thus oxygen -- abundance require a lower mass of atmospheric pollution to achieve C/O $>1$.  The prototype 
dC star G77-61 is a single-lined spectroscopic binary with a 245\,d period \citep{dea86}, and an ongoing radial velocity study of a few 
dozen dC stars is consistent with a 100\% binary fraction \citep{whi18}.  Together these two facts imply the dC stars are analogous to 
CEMP-s stars \citep{sta14,han16}.  The above facts may suggest that thin disk dC stars are rare, and only occur in systems with orbits 
favorable to efficient mass transfer, and where the carbon-rich giant evolves from a sufficiently early-type progenitor \citep{lau07}.

An intriguing possibility is that the dC stars are {\em not only the dominant type of carbon star, but also the most common CEMP 
star, in the Galaxy}.  This prospect has a firm basis in the behavior of the initial stellar mass function, at least in the disk, where far
more low-mass stars form than stars of higher mass, and specifically there are over three times as many M-type dwarfs than FGK 
stars\footnote{www.recons.org}.  The occurrence of red dwarf stars peaks somewhere between M3 and M4 and in the mass range 
$0.2-0.4$\,$M_{\odot}$, both as field stars and as companions to the intermediate-mass progenitors of white dwarfs \citep{far05,
der14}.  If correct, and with future modeling that incorporates the molecular bands that dominate dC star spectra, these cool 
dwarfs may provide the largest sample of nearby stars that represent the earliest stages of Galactic chemical evolution. The 
full potential of dC stars and their exact Galactic component fractions should soon be resolved via {\em Gaia} parallaxes.

\section*{Acknowledgements}

The authors acknowledge V. Belokurov and S. E. Koposov for providing their proper motion catalog, J. A. Munn for exchanges
on the current state of dC star and SDSS astrometry, H. C. Harris and C. A. Tout for input on an early draft.  This research has 
made use of the VizieR catalogue access tool, CDS, Strasbourg, France.

\bsp    % typesetting comment
\label{lastpage}
\end{document}